\documentclass{article}

\usepackage[dblblindworkshop, final]{neurips_2026}

\usepackage[utf8]{inputenc} 
\usepackage[T1]{fontenc}    
\usepackage{hyperref}       
\usepackage{url}            
\usepackage{booktabs}       
\usepackage{amsfonts}       
\usepackage{nicefrac}       
\usepackage{microtype}      
\usepackage{xcolor}         
\usepackage{amsmath}
\usepackage{graphicx}

\title{Reducing the Uncertainty in Determining the Origin of Natural Commodities via Earth Embeddings}
\workshoptitle{2nd Workshop on Advances in Representation Learning for Earth Observation}

\author{%
  Sebastián Jiménez$^{1}$ \quad
  Willem Waegeman$^{1}$ \quad
  Victor Deklerck$^{2}$ \quad
  Thomas Mortier$^{3}$ \\[2pt]
  \textnormal{$^{1}$Department of Data Analysis and Mathematical Modeling, Ghent University, Belgium} \\
  \textnormal{$^{2}$Meise Botanic Garden, Meise, Belgium} \quad
  \textnormal{$^{3}$Department of Environment, Ghent University, Belgium} \\[2pt]
  \texttt{\{sebastian.jimenez, willem.waegeman, thomasf.mortier\}@ugent.be} \\
  \texttt{victor.deklerck@worldforestid.org}
}

\begin{document}

\maketitle

\begin{abstract}
    Verifying where commodities like timber were harvested is a forensic task, needed to detect illegal trade due to illegal deforestation or trade sanctions. Existing Gaussian-process models of wood chemistry give a Bayesian posterior
  over candidate harvest locations, but their credible regions in general do not deliver
  the coverage they promise (0.82-0.91 at the nominal 95\%). In this work we bring in a
  second, independent source of evidence: a prior over harvest locations learned
  from satellite (AlphaEarth) embeddings. We fit one classifier per genus that predicts presence from an embedding vector, supervised by occurrence records, and normalize its scores into a prior over the candidate grid. For
  such a prior to act on the prediction sets rather than merely mask cells, we
  build a cross-conformal predictor on the \emph{posterior cell mass}, so the
  prior moves every p-value while the coverage guarantee survives any
  misspecification of it. On four European genera, the embedding prior shrinks
  regions by up to 42\% compared to a uniform prior at no cost in coverage, which holds at 0.953-0.968
  against a nominal 0.95.
\end{abstract}

\section{Introduction}
\label{sec:intro}

Since the 2022 invasion of Ukraine, sanctions have closed the Russian and Belarusian timber trade that had supplied a substantial share of the EU's wood-product imports \citep{mortier2024framework}, and the EU Deforestation Regulation now requires operators to declare the geolocation of every plot on which a commodity was produced \citep{eudr2023}. Enforcement must test such claims from the wood itself, via stable isotope ratios (SIRA) and trace element concentrations (TEA), which vary predictably across space \citep{dormontt2015forensic}. 

\citet{mortier2024framework} combine Gaussian processes per chemical marker, whose likelihoods are then multiplied and used in a Bayesian inference step to find a provenance probability over candidate harvest locations. Related pipelines to this approach are already deployed with European enforcement agencies \citep{sarkar2025chasing}. We ask how trustworthy the predictive region is: the chemistry localises a sample only to 180-230\,km from the true origin on average \citet{mortier2024framework}, and its nominal-95\% credible regions cover the truth 0.82-0.91 of the time in our replication (0.60 for \textit{Pinus}). 

We therefore bring a second source of evidence that is independent of the wood chemistry: the AlphaEarth Foundations embeddings \citep{brown2025alphaearth}, which summarise multi-sensor optical and radar time series into a 64-dimensional vector per candidate location. Because these vectors already encode vegetation, terrain, and climate context, a simple $L_2$-regularised logistic regression on the embedding recovers per-genus presence across Europe, without the need for bioclimatic covariates. We fit one such classifier per genus, restricted to forested land so that the prior places no mass on cells no timber could have come from, and use its scores as a prior over candidate harvest locations. We then quantify the impact on coverage and region size of predictions compared to a non-informative uniform prior.

To validate the improvement obtained with the earth embeddings as the prior source of information, we use conformal prediction \citep{vovk2005algorithmic,angelopoulos2021gentle}, which guarantees the probabilistic credible regions validity under exchangeability alone, for any conformity score, whether or not the Gaussian processes, the
independence assumption, or the prior are correct, reaching a provable finite-sample floor coverage \citep{barber2021predictive}. We
rebuild the predictor around the negative log posterior \emph{cell mass}, a
score in which the prior reranks candidates and moves every p-value, over a
candidate set restricted only by the genus's chorological range; this is what
distinguishes the problem from Bayes-assisted conformal regression
\citep{fong2021conformal,deliu2025interplay}, from conformal species
distribution models \citep{poisot2024conformal} and from spatial conformal
prediction over continuous responses \citep{mao2024valid}. Evaluated with a
nested protocol over four genera, two priors and four seeds, the Earth-embedding prior shrinks regions by up to 42\% (\textit{Quercus},
$\alpha = 0.1$) at no cost in coverage, which sits at 0.953-0.968 at
$\alpha = 0.05$ against a nominal 0.95; the Bayesian regions offer no such
guarantee, provides an empirical coverage of 0.82-0.91.

\section{Method}
\label{sec:method}
\textbf{Setting.} Given a reference dataset $\mathcal{D}=\{(x_i, y_i)\}_{i=1}^n$, where $x_i \in
\mathbb{R}^m$ collects the isotope ratios and trace element concentrations of
sample $i$ and $y_i \in \mathbb{R}^2$ is its harvest location, we model each
measured chemical variable $t \in \{1,\dots,m\}$ independently as
\begin{equation}
  x^{(t)} = f^{(t)}(y) + \epsilon^{(t)}, \qquad
  \epsilon^{(t)} \sim \mathcal{N}\!\bigl(0, \sigma^{2\,(t)}_m\bigr), \qquad
  f^{(t)}(y) \sim \mathcal{GP}\bigl(c^{(t)},\, k^{(t)}(y)\bigr),
\end{equation}
with a constant mean $c$ and a Mat\'ern kernel $k$ and noise variance $\sigma^2_m$,  following the determination model
of \citet{mortier2024framework}. Fitted on a training set, each GP yields
predictive moments $\hat{\mu}^{(t)}(y)$, $\hat{\sigma}^{2\,(t)}(y)$ at any
location $y$, and the likelihood of a measurement vector $x$ factorises as
$p(x \mid y) = \prod_{t=1}^m \phi\bigl(x^{(t)} \mid \hat{\mu}^{(t)}(y),
\hat{\sigma}^{2\,(t)}(y)\bigr)$.

\textbf{Candidate set and priors.}
For inference, we use candidate locations that form a grid $G$ of $0.2^\circ$ cells over the study area, restricted to the genus's chorological range polygon \citep{caudullo2017chorological}, an expert-drawn distribution map: $G'=G\cap R$. Every prior below therefore acts on top of an existing map.

Each cell $g\in G'$ is summarised by an embedding $z(g)\in\mathbb{R}^{64}$. obtained by averaging the AlphaEarth Foundations embeddings \citep{brown2025alphaearth} of the forested pixels within the cell at a resolution of $100$ meters, averaged across the years 2019-2024. These vectors encode vegetation, terrain and climate context learned from multi-sensor optical and radar time series. We score cells with an $L_2-$regularised logistic regression $h$ trained to separate cells where the genus is recorded in EU-Forest \citep{mauri2017euforest} from target-group background cells drawn from all forest areas in Europe.

We then take
\begin{equation}
\pi_\beta(g) \;\propto\; h\bigl(z(g)\bigr)^{\beta}, \qquad g \in G',
\label{eq:prior}
\end{equation}
normalised over $G'$ and compare $\beta = 0$ (the \emph{uniform} prior over the range map)
with $\beta = 1$ (the \emph{embedding} prior). $h$ is fitted on
occurrence records only, never on the reference samples, so no information
about any train/test split of the reference set can leak into it, and one
fitted prior serves every fold and seed. Bayes' rule gives the posterior
$p(g \mid x) \propto p(x \mid g)\, \pi_\beta(g)$, normalised over $G'$.

\textbf{Cross-conformal prediction.}
We use the negative log posterior cell probability as the nonconformity measure of an example $(x, g)$, given training data $\mathcal{D}_{train}$, 
\begin{equation}
  A\bigl(\mathcal{D}_{train}, (x, g)\bigr) \;=\; -\log p(g \mid x)
  \;=\; -\log p(x \mid g) - \log \pi_\beta(g) + \log \textstyle\sum_{g' \in
  G'} p(x \mid g')\,\pi_\beta(g'),
  \label{eq:score}
\end{equation}
where the GPs behind $p(x \mid g)$ are fitted on $\mathcal{D}_{train}$. Calibration examples are
scored at the grid cell containing their true location, the posterior being a
distribution over cells; a truth outside $G'$ receives $A = +\infty$, a rule
applied identically to calibration and test samples, so exchangeability is
preserved and the effect is conservative.
Following \citet{vovk2015cross}, we use a $4-$fold conformal prediction where scores are averaged over the models fitted on each fold. Cross-conformal prediction guarantees
a coverage of at least $1 - 2\alpha$ under
exchangeability alone \citep{vovk2015cross}, with the sharper finite-sample
floor $1 - 2\alpha - \min\{ \nicefrac{2(1-1/K)}{(\ell/K+1)},\,
\nicefrac{(1-K/\ell)}{(K+1)} \}$ of \citet{barber2021predictive} where $K$ is the number of folds and $\ell$ the size of training data.

\textbf{Evaluation.}
Scoring a reference sample under all $K$ fold models is contaminated: $K-1$ of
them were trained on it. We therefore evaluate with a \emph{nested} protocol: an outer $K$-fold split supplies test samples no model has seen; within
each outer training set an inner $K$-fold split builds the conformal predictor set. The Bayesian comparator
is the smallest credible region holding 95\% of the posterior mass under the
same prior and support.

\section{Experiments and results}
\label{sec:results}

\paragraph{Setup.}
We use the reference database of \citet{mortier2024framework}: four genera
(\textit{Betula}, \textit{Fagus}, \textit{Quercus}, \textit{Pinus}; 174-302
samples each) with the combined SIRA+TEA panel of $m = 20$ variables, over an
Eastern-European study grid of 18{,}876 cells ($|G'|$ from 2{,}939 for
\textit{Fagus} to 14{,}675 for \textit{Betula}; at least 99.5\% of candidate
cells are scorable by the AlphaEarth embedding prior). All results use the nested
protocol and are repeated over four outer/inner cross-validation seeds; tables
report across-seed mean $\pm$ sd, and every number is generated by the
analysis code.\footnote{Code and result tables: \href{https://github.com/sebasjimenez9808/workshop-neurips-2026.git}{GitHub repository.}}

\begin{table}[t]
  \caption{Coverage and efficiency of posterior-score cross-conformal regions
  (nested evaluation, SIRA+TEA, mean $\pm$ sd over four seeds). \emph{Ceiling}
  is the fraction of test samples whose true location the candidate set $G'$
  admits at all; coverage cannot exceed it. Areas in $10^3\,$km$^2$.}
  \label{tab:main}
  \centering
  \footnotesize
  \setlength{\tabcolsep}{4pt}
  \begin{tabular}{llc cc cc }
    \toprule
    & & & \multicolumn{2}{c}{$\alpha = 0.1$} & \multicolumn{2}{c}{$\alpha = 0.05$} \\
    \cmidrule(lr){4-5} \cmidrule(lr){6-7}
    Genus & Prior & Ceiling & Coverage & Area & Coverage & Area \\
    \midrule
    \textit{Betula} & uniform & 0.96 & 0.908 $\pm$ 0.010 & 1,183.8 & 0.953 $\pm$ 0.004 & 2,643.0 \\
\textit{Betula} & embedding & 0.96 & 0.914 $\pm$ 0.010 & 1,041.8 & 0.954 $\pm$ 0.003 & 2,433.5 \\
\textit{Fagus} & uniform & 1.00 & 0.930 $\pm$ 0.012 & 488.7 & 0.968 $\pm$ 0.012 & 756.2 \\
\textit{Fagus} & embedding & 1.00 & 0.933 $\pm$ 0.013 & 390.6 & 0.967 $\pm$ 0.013 & 700.4 \\
\textit{Quercus} & uniform & 0.99 & 0.916 $\pm$ 0.005 & 1,258.6 & 0.963 $\pm$ 0.002 & 2,140.5 \\
\textit{Quercus} & embedding & 0.99 & 0.915 $\pm$ 0.007 & 734.7 & 0.960 $\pm$ 0.006 & 1,664.8 \\
\textit{Pinus} & uniform & 0.66 & 0.655 $\pm$ 0.000 & 3,120.8 & 0.655 $\pm$ 0.000 & 3,120.8 \\
\textit{Pinus} & embedding & 0.66 & 0.655 $\pm$ 0.000 & 3,120.8 & 0.655 $\pm$ 0.000 & 3,120.8 \\
\bottomrule

  \end{tabular}
\end{table}

\begin{table}[t]
  \caption{Conformal regions at $\alpha = 0.025$ against nominal-95\% Bayesian credible regions
  under the same prior at the published $\nicefrac{3}{4}$ training budget
  (mean $\pm$ sd over seeds; areas in $10^3\,$km$^2$).}
  \label{tab:bayes}
  \centering
  \footnotesize
  \setlength{\tabcolsep}{4.5pt}
  \begin{tabular}{ll cc cc c}
    \toprule
    & & \multicolumn{2}{c}{Conformal ($\alpha=0.025$)} & \multicolumn{2}{c}{Bayes 95\%} & \\
    \cmidrule(lr){3-4} \cmidrule(lr){5-6}
    Genus & Prior & Coverage & Area & Coverage & Area & Ratio \\
    \midrule
    \textit{Betula} & embedding & 0.964 $\pm$ 0.000 & 3,875.6 & 0.821 $\pm$ 0.013 & 465.5 & 8.33 \\
\textit{Fagus} & embedding & 0.982 $\pm$ 0.009 & 846.7 & 0.908 $\pm$ 0.017 & 320.7 & 2.64 \\
\textit{Quercus} & embedding & 0.983 $\pm$ 0.003 & 2,666.1 & 0.889 $\pm$ 0.008 & 497.5 & 5.36 \\
\textit{Pinus} & embedding & 0.655 $\pm$ 0.000 & 3,120.8 & 0.599 $\pm$ 0.002 & 452.4 & 6.90 \\
\bottomrule

  \end{tabular}
\end{table}

\begin{figure}[t]
  \centering
  \includegraphics[width=0.88\linewidth]{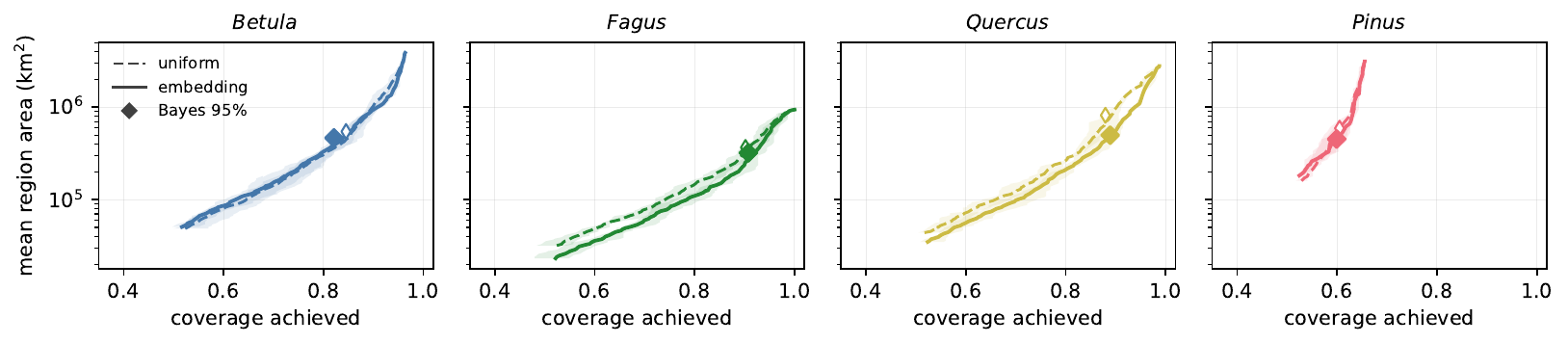}
  \caption{Mean region area (log scale) against achieved coverage, uniform
  (dashed) vs.\ Earth embedding (solid) prior; bands span the four seeds. Diamonds:
  nominal-95\% Bayesian regions at their \emph{achieved} coverage (filled:
  Earth embedding; open: uniform).}
  \label{fig:area}
\end{figure}

\begin{figure}[t]
  \centering
  \includegraphics[width=0.88\linewidth]{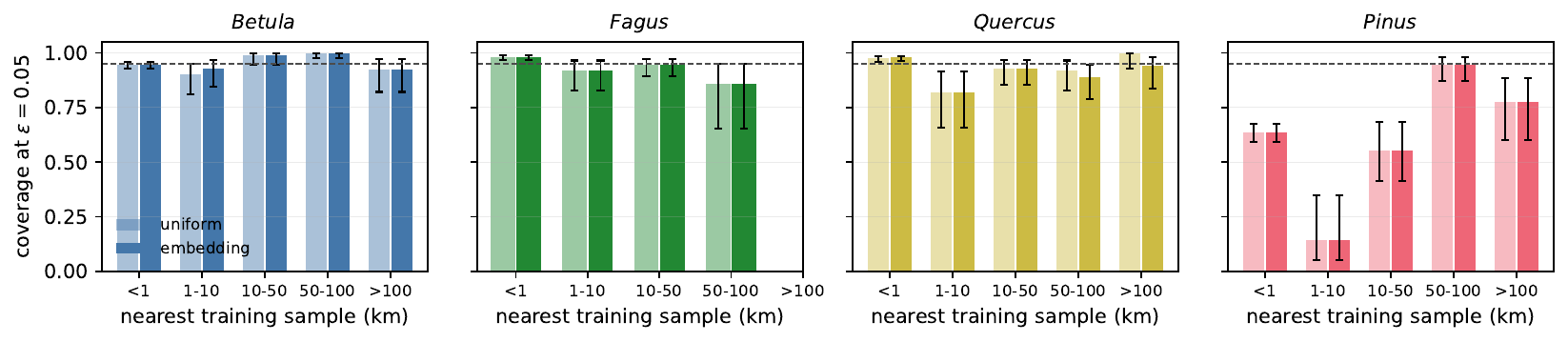}
  \caption{Coverage at $\alpha = 0.05$ conditioned by distance to the nearest training
  sample (pooled across seeds; Wilson 95\% intervals). Most samples have a
  near-duplicate in training; the right-hand strata stress-test the
  guarantee.}
  \label{fig:remoteness}
\end{figure}

\textbf{Validity.}
Table~\ref{tab:main} shows the empirical coverage sits at or
just above the nominal $1-\alpha$ (0.953--0.968 at $\alpha = 0.05$),
comfortably clearing the $1-2\alpha$ guarantee and the finite-sample floor
of \citet{barber2021predictive}. For \textit{Pinus} the ceiling binds: 34.5\% of true locations fall outside $G'$, coverage is pinned at 0.655, making its
regions conservative as well as capped.

\textbf{Does the informative prior help?}
Yes, at no cost in coverage, and by an amount that varies by genus
(Table~\ref{tab:main}, Figure~\ref{fig:area}). At $\alpha = 0.1$ the
Earth embedding prior shrinks mean region area by 42\% for \textit{Quercus}, 20\% for \textit{Fagus} and 12\% for
\textit{Betula}; at $\alpha = 0.05$ the gains compress to 22\%, 7\% and
8\%. \textit{Pinus} is unchanged, its regions are already pinned
near the whole candidate set by the ceiling. The reasons for the differences in impact can come from a biological perspective (different genera can grow in more diverse soils), the structure of the dataset (some genus samples are more concentrated than others in more homogeneous spaces), or the explainability power of the Earth embeddings (the variance of each dimension can not be enough to explain the differences between candidate locations).

\textbf{Conformal vs.\ Bayes, and robustness.}
Table~\ref{tab:bayes} shows that the
nominal-95\% credible regions are far smaller but deliver only 0.82-0.91
coverage on held-out samples (0.60 for \textit{Pinus}), while the conformal
regions at $\alpha = 0.025$ carry a provable ${\geq}92\%$ guarantee
(empirically 0.964-0.983) at 2.6-8.3$\times$ their area. At \emph{matched achieved} coverage the comparison inverts; and for the wide-range genera the guaranteed regions
saturate (\textit{Betula} covers 0.964 with a region close to all of $G'$). 

\textbf{Robustness.}
Across the four cross-validation seeds, coverage at $\alpha = 0.05$ moves by
at most 0.027 and the area ranking between priors never flips. Stratified by distance to the nearest training sample
(Figure~\ref{fig:remoteness}), coverage holds in the ${<}1$\,km stratum that
contains roughly 70\% of samples and in most others; the sparse middle strata
dip (e.g.\ \textit{Quercus} $0.82$ at 1-10\,km, $n = 33$) with intervals wide
enough that no stratum falls significantly below nominal except
\textit{Pinus}, whose stratified pattern traces which regions its range map
excludes rather than any failure of the calibration. The stability of the coverage conditioned on distance to the closest training sample shows that spatial dependency might not be a decisive factor in the validity of the conformal prediction step.

\section{Conclusion}
\label{sec:conclusion}

Currently used origin identification models do not deliver
the empirical coverage their nominal confidence advertises. We show that a
prior built from AlphaEarth embeddings, which encode environmental conditions
globally, can be combined with the existing chemistry model inside a
cross-conformal predictor, yielding regions whose coverage is guaranteed. First, wherever the
candidate set contains the true origin, the regions are calibrated almost
exactly at $1-\alpha$. Second, the Earth embedding prior, fitted
purely on occurrence data and hence leakage-free by construction, generates up to 42\% smaller regions at
$\alpha = 0.1$, at no cost in coverage. Third, the Bayesian credible regions
are smaller than the conformal ones but reach only 0.82--0.91 of their
nominal 95\% on held-out samples; at matched achieved coverage the gap
largely disappears. This work has two main limitations. Firstly, the potential spatial dependence of the data challenges the exchangeability assumption of the conformal prediction framework. To test the robustness of the method and the potential impact of this, Figure~\ref{fig:remoteness} shows the coverage for samples that are more than $100$ kilometers away from the closest training sample. This represents a region that was left out during calibration. Future work will discuss this in more detail and present alternatives for a more systematic evaluation of the exchangeability assumption. Secondly, the mechanisms of why the informative prior effect varies across genera are not clear. This analysis would require a closer look at the biological aspects of the forests where each genus can grow and the variance of the embeddings' dimensions across each genus range.

\section*{Acknowledgments}
S.J. was supported by the Research Foundation Flanders (FWO Vlaanderen) (PhD fellowship, fundamental research grant 11A3G26N). W.W. received funding from the Flemish government under the ``Flanders AI Research" program. 

\bibliographystyle{plainnat}
\bibliography{refs}

\medskip





\end{document}